\documentclass[aps,pra,twocolumn,groupedaddress,showpacs]{revtex4}
\usepackage{amsfonts}
\usepackage{amssymb}
\usepackage{bm}
\usepackage{graphicx}%
\usepackage{hyperref}
\usepackage{amsmath}
\newcommand{\ket}[1]{\ensuremath{\left|{#1}\right\rangle}}
\newcommand{\vect}[1]{\ensuremath{\bm{#1}}}
\newcommand{\bra}[1]{\ensuremath{\left\langle{#1}\right |}}
\newcommand{\oper}[1]{\bm{\mathsf{#1}}}
\newcommand{\bsy}[1]{\ensuremath{\boldsymbol{#1}}}
\newcommand{\brm}[1]{\ensuremath{\mathbf{#1}}}

\newcommand{\sinc}[1]{\ensuremath{\mathrm{sinc}\, #1}}

\begin{document}
\title{Entanglement and conservation of orbital angular momentum in spontaneous parametric down-conversion}
\author{S. P. Walborn}
\affiliation{Universidade Federal de Minas Gerais, Caixa Postal 702, Belo Horizonte, MG 30123-970, Brazil}
\author{A. N. de Oliveira}
\affiliation{Universidade Federal de Minas Gerais, Caixa Postal 702, Belo Horizonte, MG 30123-970, Brazil}
\author{R. S. Thebaldi}
\affiliation{Universidade Federal de Minas Gerais, Caixa Postal 702, Belo Horizonte, MG 30123-970, Brazil}
\author{C. H. Monken}
\email[]{monken@fisica.ufmg.br}
\affiliation{Universidade Federal de Minas Gerais, Caixa Postal 702, Belo Horizonte, MG 30123-970, Brazil}

\date{\today}

\begin{abstract}
We show that the transfer of the plane wave spectrum of the pump beam to the fourth-order transverse spatial correlation function of the two-photon field generated by spontaneous parametric down-conversion leads to the conservation and entanglement of orbital angular momentum of light.  By means of a simple experimental setup based on fourth-order (or two-photon) interferometry, we show that our theoretical model provides a good description for down-converted fields carrying orbital angular momentum.  
\end{abstract}


\maketitle

\section{Introduction}
It is well established that when the paraxial approximation is valid, any electromagnetic beam with an azimuthal phase dependence of the form $e^{il\phi}$ carries an orbital angular momentum $l\hbar$ per photon \cite{allen92}. Interesting enough by itself due to its fundamental character, this fact also raises possibilities for technical applications. For example, in the rapidly developing field of quantum information, it has been pointed out recently that it is possible to increase the amount of information carried by a single photon by encoding qubits in the orbital angular momentum \cite{arnaut00,eliel00}. Laguerre-Gaussian (LG hereafter) beams are the most known and studied examples of beams carrying orbital angular momentum.  
Devices that discriminate the orbital angular momentum of Laguerre-Gaussian  beams have been reported and experimentally tested for very low intensities, suggesting that they should work at the single photon level \cite{leach02}. 

An important potential application of light beams carrying orbital angular momentum is the generation of photon pairs with discrete multidimensional entanglement \cite{mair01}. This can be obtained by means of spontaneous parametric down-conversion (SPDC) pumped by a LG beam. Denoting by $\ket{m}$ a one-photon state carrying an orbital angular momentum $m\hbar$ and by $l$ the azimuthal index of the the LG pump beam, the two-photon state generated by SPDC can be written as
\begin{equation}
\ket{\psi}=\sum_{m=-\infty}^{+\infty}C_{m}\ket{l-m}\ket{m}.
\end{equation}
This expression is based on the hypothesis that orbital angular momentum is conserved  in SPDC. Some authors have studied this issue \cite{arnaut00,arnold02,barbosa02,torres03}, and experimental results \cite{mair01} suggest that orbital angular momentum is in fact conserved. One has to consider, however, that although the process of down-conversion itself may conserve angular momentum, in most cases, the pump beam propagates in a birefringent nonlinear crystal as an extraordinary beam. The anisotropy of the medium causes a small astigmatism in the LG beam as it propagates, breaking its circular symmetry in the transverse plane. This  symmetry breaking is equivalent to an exchange of angular momentum between the medium and the pump beam, so that the conservation holds only on average. This effect depends on both the angular momentum of the pump beam and on the crystal length, being negligible for thin crystals and low values of $l$. A detailed account of this problem will be published elsewhere.

In an earlier paper \cite{monken98a}, we showed that the phase matching conditions in SPDC are responsible for a transfer of the amplitude and phase characteristics of the pump beam to the two-photon field. In fact, it is the plane wave spectrum, or the so-called angular spectrum of the pump beam that is transfered to the fourth-order spatial correlation properties of the down-converted field. In this work, we demonstrate theoretically and experimentally that the conservation of orbital angular momentum as well as the multidimensional entanglement in the SPDC process in the thin crystal paraxial approximation is a direct consequence of the transfer of the plane wave spectrum from the pump beam to the two-photon state. By means of a simple experimental setup based on fourth-order (or two-photon) interferometry, we show that our theoretical model provides a good description for down-converted fields carrying orbital angular momentum.  

\section{Theory}
\subsection{The state generated by SPDC}  
In the monochromatic and paraxial
approximations, the two-photon quantum state generated by
non-collinear SPDC can be written as \cite{hong85,monken98a}
\begin{equation} 
\ket{\psi} = C_{1}\ket{\mathrm{vac}} + C_{2}\ket{\psi}
\end{equation} 
where 
\begin{equation}
\ket{\psi}=\sum_{\sigma_{s},\sigma_{i}}C_{\sigma_{s},\sigma_{i}}\int\hspace{-2mm}\int\limits_{D}\hspace{-1mm} d\brm{q}_{s}
d\brm{q}_{i}\ \Phi(\brm{q}_{s},\brm{q}_{i})\ket{\brm{q}_{s},\sigma_{s}}_{s}
\ket{\brm{q}_{i},\sigma_{i}}_{i}. 
\label{eq:state} 
\end{equation} 
The coefficients $C_1$ and $C_2$ are such that $|C_{2}| \ll \, |C_{1}|$. $C_2$ depends on the crystal length, the nonlinearity coefficient, the magnitude of the pump field, among other factors. The kets $\ket{\brm{q}_{j},\sigma_{j}}$ represent one-photon states in plane wave modes labeled by the transverse component $\brm{q}_{j}$ of the wave vector
$\brm{k}_{j}$ and by the polarization $\sigma_{j}$ of the mode $j = s$ or $i$. The polarization state of the down-converted photon pair is defined by the coefficients $C_{\sigma_{s},\sigma_{i}}$. The function $\Phi(\brm{q}_{s},\brm{q}_{i})$ is given by \cite{monken98a}
\begin{equation} 
\Phi(\brm{q}_{s},\brm{q}_{i}) =\frac{1}{\pi}\sqrt{\frac{2L}{K}}\ 
v(\brm{q}_{s}+\brm{q}_{i})\ 
\sinc\left(\frac{L|\brm{q}_{s}-\brm{q}_{i}|^{2}}{4K} \right), 
\label{eq:state2}
\end{equation} 
where $v(\brm{q})$ is the normalized angular spectrum of the pump beam, $L$ is the length of the nonlinear crystal in the propagation ($z$) direction, and $K$ is the magnitude of the pump field wave vector. The integration domain $D$ is, in
principle, defined by the conditions $q_{s}^{2}\le k_{s}^{2}$ and $q_{i}^{2}\le k_{i}^{2}$. However, in most experimental conditions, the domain in which $\Phi(\brm{q}_{s},\brm{q}_{i})$ is appreciable is much smaller.  If the crystal is thin enough, the sinc function in (\ref{eq:state}) can be approximated by 1. We assume that $\Phi(\brm{q}_{s},\brm{q}_{i})$ does not depend on the polarizations of the down-converted photons. In some cases, this is not true, especially when one is dealing with type-II phase matching, in which case the two photons have orthogonal polarizations. However, this dependence can be made negligible by the use of compensators in the down-converted beams \cite{kwiat95}.

The two-photon detection amplitude, which can be regarded as a photonic
wave function is
\begin{equation}
\bsy{\Psi}(\brm{r}_s,\brm{r}_i) =
\bra{\mathrm{vac}}\brm{E}_{i}^{(+)}(\brm{r}_i)\brm{E}_{s}^{(+)}(\brm{r}_s)\ket{\psi},
\label{eq:wf}
\end{equation}
where $\brm{E}_{j}^{(+)}(\brm{r})$ is the field operator for the plane wave mode $j$. In the paraxial approximation, $\brm{E}_{j}^{(+)}(\brm{r})$ is 
\begin{equation}
\brm{E}_{j}^{(+)}(\brm{r}) = e^{ikz} \sum\limits_{\sigma}\int d\brm{q}
\,\oper{a}_{j}(\brm{q},\sigma)\vect{\epsilon}_{\sigma}e^{i(\brm{q} \cdot
\vect{\rho}-\frac{q^{2}}{2k}z)}.
\label{eq:field}
\end{equation}
The operator $\oper{a}_{j}(\brm{q},\sigma)$ annihilates a photon in mode $j$ with transverse wave vector $\brm{q}$ and polarization $\sigma$.

In the analysis that follows, we do not need to consider polarization. So, $\bsy{\Psi}(\brm{r}_{s},\brm{r}_{i})$ will be treated as a scalar function. 
In addition, we will work in the far field and make the following simplifications: $z_{s}=z_{i}=Z$, $k_{s}=k_{i}=\frac{1}{2}K$.
It is known that if the paraxial approximation is valid, the two-photon wave function is
\begin{equation}
\Psi(\bsy{\rho}_{s},\bsy{\rho}_{i},z_{s},z_{i}) = \mathcal{E}\left(\frac{\bsy{\rho}_{s}+\bsy{\rho}_{i}}{2},Z\right)\ \mathcal{F}\left(\frac{\bsy{\rho}_{s}-\bsy{\rho}_{i}}{\sqrt{2}},Z\right),
\label{eq:psif}
\end{equation} 
where $\mathcal{E}(\bsy{\rho},z)$ is the normalized electric field amplitude of the pump beam and $\mathcal{F}(\bsy{\rho},z)=\frac{\sqrt{KL}}{2\pi z}\ \sinc(\frac{KL}{8z^{2}}\rho^{2})$. In order to clean-up the notation, we will omit the dependence on the $z$ coordinate hereafter. We see that the two-photon wave function $\Psi$ carries the same functional form as the pump beam amplitude, calculated in the coordinate $\bsy{\rho}=\frac{1}{2}\bsy{\rho}_{s}+\frac{1}{2}\bsy{\rho}_{i}$. The pump beam field $\mathcal{E}(\bsy{\rho})$ is characterized by its wavelength $\lambda_{o}$ and its waist $w_{o}$. To be more precise, we will write $\mathcal{E}$ as  $\mathcal{E}(\bsy{\rho};\lambda_{o},w_{o})$. Since we are working with down-converted fields satisfying $\lambda_{s}=\lambda_{i}=2\lambda_{o}$, it is convenient to write $\Psi$ in terms of a beam with the same angular spectrum of the pump field, as required by Eq. (\ref{eq:psif}), but with a wavelength $\lambda_{c}=2\lambda_{o}$, and a waist $w_{c}=\sqrt{2}\,w_{o}$. From the general form of gaussian beams, apart from a phase factor and normalization constants, it is evident that
\begin{equation}
\mathcal{E}(\bsy{\rho};\lambda_{o},w_{o})=\mathcal{E}(\sqrt{2}\,\bsy{\rho};2\lambda_{o},\sqrt{2}\,w_{o})\equiv \mathcal{U}(\sqrt{2}\,\bsy{\rho}).
\end{equation}
So, $\Psi$ can be put in the more convenient form
\begin{equation}
\Psi(\bsy{\rho}_{s},\bsy{\rho}_{i}) = \mathcal{U}\left(\frac{\bsy{\rho}_{s}+\bsy{\rho}_{i}}{\sqrt{2}}\right)\ \mathcal{F}\left(\frac{\bsy{\rho}_{s}-\bsy{\rho}_{i}}{\sqrt{2}}\right)
\label{eq:psif2}
\end{equation}

Let us now suppose that the down-converter is pumped by a LG beam whose orbital angular momentum is $l\, \hbar$ per photon, described by the amplitude $\mathcal{E}_{p}^{l}(\bsy{\rho};\lambda_{o},w_{o})$. Here, $p$ is the radial index. In order to study the conservation of angular momentum in SPDC, we will expand the two-photon wave function $\Psi(\bsy{\rho}_{s},\bsy{\rho}_{i})$ in terms of the LG basis functions $\mathcal{U}_{p_{s}}^{l_{s}}(\bsy{\rho}_{s})\,\mathcal{U}_{p_{i}}^{l_{i}}(\bsy{\rho}_{i})$. That is,
\begin{equation}
\Psi(\bsy{\rho}_{s},\bsy{\rho}_{i})=\sum_{l_{s},p_{s}}\sum_{l_{i},p_{i}}C_{p_{s}p_{i}}^{l_{s}l_{i}}\mathcal{U}_{p_{s}}^{l_{s}}(\bsy{\rho}_{s})\,\mathcal{U}_{p_{i}}^{l_{i}}(\bsy{\rho}_{i})
\end{equation}
From the orthogonality of the LG basis, $C_{p_{s}p_{i}}^{l_{s}l_{i}}$ is given by
\begin{align}
C_{p_{s}p_{i}}^{l_{s}l_{i}} =&\int\hspace{-2mm}\int\hspace{-1mm} d\bsy{\rho}_{s}
d\bsy{\rho}_{i}\  \mathcal{U}_{p}^{l}\left(\frac{\bsy{\rho}_{s}+\bsy{\rho}_{i}}{\sqrt{2}}\right)\,\mathcal{F}\left(\frac{\bsy{\rho}_{s}-\bsy{\rho}_{i}}{2}\right)\nonumber \\ &\times \mathcal{U}_{p_{s}}^{*l_{s}}(\bsy{\rho}_{s})\,\mathcal{U}_{p_{i}}^{*l_{i}}(\bsy{\rho}_{i}).
\label{eq:cdef}
\end{align} 

Let us make the following coordinate transformation in Eq. (\ref{eq:cdef}):
$\brm{R}=\bsy{\rho}_{s}+\bsy{\rho}_{i}$ and $\brm{S}=\frac{1}{2}(\bsy{\rho}_{s}-\bsy{\rho}_{i})$. So,
\begin{align}
C_{p_{s}p_{i}}^{l_{s}l_{i}} =&\int\hspace{-2mm}\int\hspace{-1mm} d\brm{R}\,
d\brm{S}\  \mathcal{U}_{p}^{l}\left(\frac{\brm{R}}{\sqrt{2}}\right)\,\mathcal{F}(\sqrt{2}\,\brm{S})\nonumber\\ &\times\mathcal{U}_{p_{s}}^{*l_{s}}\left(\frac{\brm{R}}{2}+\brm{S}\right)\,\mathcal{U}_{p_{i}}^{*l_{i}}\left(\frac{\brm{R}}{2}-\brm{S}\right).
\label{eq:cdef2}
\end{align} 
When $L$ is small enough (the thin crystal approximation), $\mathcal{F}$ can be approximated by 1 in Eq. (\ref{eq:cdef2}), provided the order of the LG modes ($N=2p+|l|$) is not too large.  In this case, the integral in $\brm{S}$ is proportional to $\mathcal{U}_{p_{s}}^{*l_{s}}(\brm{R})*\mathcal{U}_{p_{i}}^{*l_{i}}(\brm{R})$, that is, the convolution of $\mathcal{U}_{p_{s}}^{*l_{s}}$ and $\mathcal{U}_{p_{i}}^{*l_{i}}$. Numerical calculations show that in the worst case, that is, $p_{s}=p_{i}$, $|l_{s}|=|l_{i}|$ and $R=0$, for a 1 mm-thick crystal, pumped by a laser with $\lambda=351$ nm and a waist of $w_{0}=1$ mm, the mean square error is less than 1\% for $N=100$. Since we are neglecting the effects due to the anisotropy of the crystal, as discussed before, there is no point in seeking exact solutions for large values of $N$. 

Under the thin crystal approximation, Eq. (\ref{eq:cdef2}) is more conveniently written in terms of Fourier transforms, as
\begin{align}
C_{p_{s}p_{i}}^{l_{s}l_{i}}\propto &\int\hspace{-2mm}\int\hspace{-2mm}\int\hspace{-1mm} d\brm{R}\,
d\brm{q}\,d\brm{q}'\   \mathcal{V}_{p}^{l}(\sqrt{2}\,\brm{q}')\nonumber\\ &\times\mathcal{V}_{p_{s}}^{*l_{s}}(\brm{q})\,\mathcal{V}_{p_{i}}^{*l_{i}}(\brm{q})\ e^{i\brm{R}\cdot(\brm{q}'-\brm{q})} \nonumber \\
\propto &\int\hspace{-1mm} d\brm{q}\   \mathcal{V}_{p}^{l}(\sqrt{2}\,\brm{q})\,\mathcal{V}_{p_{s}}^{*l_{s}}(\brm{q})\,\mathcal{V}_{p_{i}}^{*l_{i}}(\brm{q}),
\label{eq:cfour}
\end{align} 
where $\mathcal{V}_{\mu}^{\nu}(\brm{q})$ is the Fourier transform of $\mathcal{U}_{\mu}^{\nu}(\brm{R})$. Writing Eq. (\ref{eq:cfour}) in cylindrical coordinates $\brm{q}\rightarrow (q,\phi)$, the LG profiles are $\mathcal{V}_{\mu}^{\nu}(\brm{q}) = v_{\mu}^{\nu}(q)\, e^{i\nu\phi}$. Then, we have
\begin{equation}
C_{p_{s}p_{i}}^{l_{s}l_{i}}\propto \int\hspace{-2mm}\int\hspace{-1mm} 
q\,dq\,d\phi \   v_{p}^{l}(\sqrt{2}\,q)\,v_{p_{s}}^{*l_{s}}(q)\,v_{p_{i}}^{*l_{i}}(q)\ e^{-i(l_{s}+l_{i}-l)\phi},
\end{equation}
that is,
\begin{equation}
C_{p_{s}p_{i}}^{l_{s}l_{i}}\propto \delta_{l_{s}+l_{i},l}\int\hspace{-1mm} q\,dq\   v_{p}^{l}(\sqrt{2}\,q)\,v_{p_{s}}^{*l_{s}}(q)\,v_{p_{i}}^{*l_{i}}(q).
\label{eq:cfour2}
\end{equation} 

Thus, orbital angular momentum is conserved in the SPDC process.  In principle, this conservation could be satisfied by fields exhibiting either a classical or quantum correlation (entanglement) of orbital angular momentum.  We will now show that the conservation leads to entanglement of orbital angular momentum of the down-converted fields.
\par
From (\ref{eq:psif2}) it is evident that, when $\mathcal{F} = 1$, the biphoton wave function reproduces the pump beam transverse profile.  Let us assume that (\ref{eq:psif2}) (with $\mathcal{F}=1$) accurately describes the two-photon state from SPDC and that the pump beam is a LG mode with $l \neq 0$.  Then, the biphoton wave function is
\begin{equation}
\Psi(\bsy{\rho}_{s},\bsy{\rho}_{i}) = \mathcal{U}_{p}^{l}\left(\frac{\bsy{\rho}_{s}+\bsy{\rho}_{i}}{\sqrt{2}}\right)
\label{eq:new1}
\end{equation}  
from which, it is 
evident that $\Psi(\bsy{\rho}_{s}+\bsy{\Delta},\bsy{\rho}_{i}-\bsy{\Delta})=\Psi(\bsy{\rho}_{s},\bsy{\rho}_{i})$.
Due to the phase structure of $\mathcal{U}_{p}^{l}$, for  $l \neq 0$ there exist transverse spatial positions $\bsy{\rho}_{s0}$ and $\bsy{\rho}_{i0}$ such that $\mathcal{U}_{p}^{l}(\bsy{\rho}_{s0}+\bsy{\rho}_{i0})=0$.  Then, clearly 
\begin{equation}
\Psi(\bsy{\rho}_{s0}+\bsy{\Delta},\bsy{\rho}_{i0}-\bsy{\Delta})=\Psi(\bsy{\rho}_{s0},\bsy{\rho}_{i0}) = 0
\end{equation}
and the coincidence detection probability $\mathcal{P}(\bsy{\rho}_{s},\bsy{\rho}_{i}) = |\Psi(\bsy{\rho}_{s},\bsy{\rho}_{i})|^{2}$ satisfies 
 \begin{equation}
 \mathcal{P}(\bsy{\rho}_{s0}+\bsy{\Delta},\bsy{\rho}_{i0}-\bsy{\Delta})=\mathcal{P}(\bsy{\rho}_{s0},\bsy{\rho}_{i0}) = 0
 \label{eq:new2}
 \end{equation} 
 \par
Now suppose that the down-converted fields exhibit a classical correlation that conserves orbital angular momentum.  The detection probability $\mathcal{P}_{\mathrm{cc}}$ for such a correlation can be written as
\begin{equation}
\mathcal{P}_{\mathrm{cc}}(\bsy{\rho}_{s},\bsy{\rho}_{i}) = \sum\limits_{l_{i}=-\infty}^{\infty}P_{l_{i}}|F_{l-l_{i}}(\bsy{\rho}_{s})|^{2}|G_{l_{i}}(\bsy{\rho}_{i})|^{2}
\label{eq:new3}
\end{equation}
where $F_{l_{s}}(\bsy{\rho}_{s})$ and $G_{l_{i}}(\bsy{\rho}_{i})$ represent down-converted signal and idler fields with orbital angular momentum 
$l_{s} \hbar$ and $l_{i} \hbar$ per photon, respectively.  Here  the coefficients $P_{l_{i}}$ satisfy $\sum_{l_{i}=-\infty}^{\infty}P_{l_{i}} = 1$ and $P_{l_{i}} \geq 0$.  Now, if (\ref{eq:new3}) accurately describes the two-photon state, then it must also satisfy the equivalent of (\ref{eq:new2}):
 \begin{equation}
 \mathcal{P}_{\mathrm{cc}}(\bsy{\rho}_{s0}+\bsy{\Delta},\bsy{\rho}_{i0}-\bsy{\Delta})  =\mathcal{P}_{\mathrm{cc}}(\bsy{\rho}_{s0},\bsy{\rho}_{i0})  = 0 \nonumber 
 \end{equation}
 which gives 
 \begin{equation}
\sum\limits_{l_{i}=-\infty}^{\infty}P_{l_{i}}|F_{l-l_{i}}(\bsy{\rho}_{s0}+\bsy{\Delta})|^{2}|G_{l_{i}}(\bsy{\rho}_{i0}-\bsy{\Delta})|^{2} =0 
 \label{eq:new4}
 \end{equation}  
 Since $P_{l_{i}} \geq 0$, a non-trivial solution to (\ref{eq:new4}) exists (for the cases where at least one $P_{l_{i}}\neq 0$) only if $|F_{l-l_{i}}(\bsy{\rho}_{s0}+\bsy{\Delta})|^{2} = 0$ or $|G_{l_{i}}(\bsy{\rho}_{i0}-\bsy{\Delta})|^{2} = 0$ for all $\bsy{\Delta}$, which implies that $F_{l-l_{i}}\equiv 0$ or $G_{l_{i}}\equiv 0$.  Thus, a classical
correlation of orbital angular momentum  states cannot
reproduce the two photon wave function (\ref{eq:psif2}).

With the reasoning above, we have shown that, assuming Eq. (\ref{eq:psif2}) accurately describes the biphoton wave function from SPDC, the conservation of orbital angular momentum in SPDC  is not satisfied by a classical correlation of the down-converted fields.  This implies that the fields are entangled in orbital angular momentum.

\subsection{The Hong-Ou-Mandel interferometer}
\begin{figure}
\includegraphics[width=5cm]{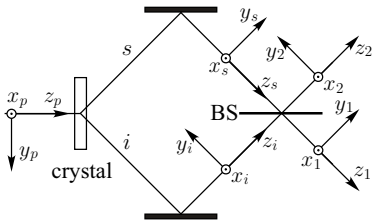}% 
\caption{The Hong-Ou-Mandel interferometer. Reflection at the beam splitter
causes a sign change in the $y$-coordinate.} 
\label{hom} 
\end{figure}

Having demonstrated that the two-photon wave function (\ref{eq:psif}) leads to conservation and entanglement of orbital angular momentum, the next step is to prove that it describes accurately the state generated by SPDC within the assumed approximations. Although  direct coincidence detection provides information about the modulus of $\Psi(\bsy{\rho}_{s},\bsy{\rho}_{i})$, its phase structure can only be revealed by some sort of interference measurement.   We do this with the help of the Hong-Ou-Mandel (HOM) interferometer \cite{hong87}, represented in Fig. \ref{hom} and described below. Coincidence measurements are taken from the two output ports of the beam splitter.  When the interferometer is balanced, that is, when paths $s$ and $i$ are equal, we have fourth-order interference. When the path length difference is much greater than the coherence length of the down-converted fields, the interferometer plays essentially no role other than decreasing the coincidence counts by a factor of $1/2$, and we can perform simple coincidence measurements.

In the HOM interferometer, the state (\ref{eq:state}) is incident on a symmetric beam splitter as shown in Fig. \ref{hom}. The annihilation operators in modes $1$ and $2$ after the beam splitter can be expressed in terms of the operators in modes $s$ and $i$:
\begin{align}
\oper{a}_{1}(\brm{q})& = t\oper{a}_{s}(q_{x},q_{y}) + i r
\oper{a}_{i}(q_{x},-q_{y}) \label{eq:aa} \\
\oper{a}_{2}(\brm{q})& = t\oper{a}_{i}(q_{x},q_{y}) + i r
\oper{a}_{s}(q_{x},-q_{y}),
\label{eq:ab} 
\end{align} 
where $t$ and $r$ are the transmission and reflection coefficients of the beam splitter.  The negative sign that appears in $q_{y}$ components is due to the
reflection from the beam splitter, as shown in Fig. \ref{hom}.
%---------------------------FIGURE------------------------------------- 
%--------------------------END----FIGURE-------------------------------
If $\brm{r}_{1}$ and $\brm{r}_{2}$ are the positions of detectors $D_{1}$ and $D_{2}$, each located at one output of the beam splitter, the coincidence detection amplitude is given by
\begin{equation}
\label{eq:refl}
\Psi_{c} = 
\Psi_{tt}(\brm{r}_{1},\brm{r}_{2}) +
\Psi_{rr}(\brm{r}_{1},\brm{r}_{2}),
\end{equation}
where the indices $tt$ and $rr$ refer to the cases when both photons are transmitted or reflected by the beam splitter, respectively. Combining Eq. (\ref{eq:psif2}) with $\mathcal{F}\equiv 1$ and Eqs.  (\ref{eq:aa} -- \ref{eq:refl}), it is straightforward to show that, for $t=r=1/\sqrt{2}$, apart from a common factor,
\begin{align}
\Psi_{c}(\bsy{\rho}_{1},\bsy{\rho}_{2})\propto&
\frac{1}{2}\left[ \mathcal{U}\left(\frac{x_{1}+x_{2}}{\sqrt{2}},\frac{y_{1}+y_{2}}{\sqrt{2}}\right)\right.\nonumber\\
&\left.-\mathcal{U}\left(\frac{x_{1}+x_{2}}{\sqrt{2}},\frac{-y_{1}-y_{2}}{\sqrt{2}}\right)\right].
\label{eq:psic}
\end{align}

Since the pump beam is a LG beam, $\mathcal{U}$ has the form
\begin{equation}
\mathcal{U}(\bsy{\rho})= u_{p}^{l}(\rho)e^{il\phi}.
\end{equation} 
According to Eq. (\ref{eq:psic}), the corresponding coincidence detection amplitude is
\begin{equation}
\Psi_{c}(\bsy{\rho}_{1},\bsy{\rho}_{2})=\Psi_{c}(R,\theta)\propto\, u_{p}^{l}(R)\sin l\theta,
\label{eq:ampli}
\end{equation}
where $R=\frac{1}{\sqrt{2}}|\bsy{\rho}_{1}+\bsy{\rho}_{2}|$ and $\theta$  is defined by the relations
\begin{align}
\sin{\theta}=\frac{\rho_{1}\sin{\phi_{1}}+\rho_{2}\sin{\phi_{2}}}{R}\\
\cos{\theta}=\frac{\rho_{1}\cos{\phi_{1}}+\rho_{2}\cos{\phi_{2}}}{R}
\end{align}
The coincidence detection probability, which is proportional to $|\Psi_{c}(R,\theta)|^{2}$, is
\begin{equation}
P_{12}(\bsy{\rho}_{1},\bsy{\rho}_{2})\propto |u_{p}^{l}(R)|^{2}\sin^{2} l\theta.
\label{eq:prob1}
\end{equation}

\section{Experiment}
%---------------------------FIGURE----------------------------------- 
\begin{figure}
 \includegraphics[width=8cm]{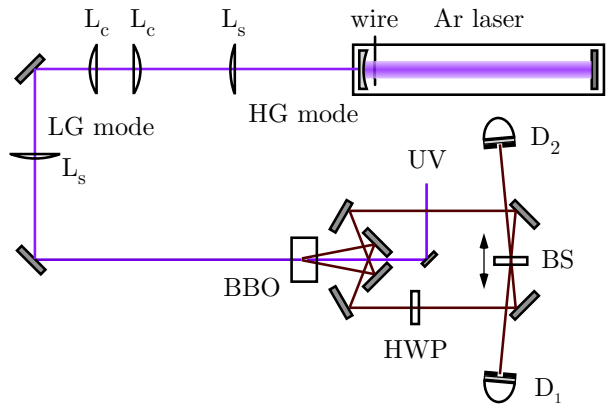}%
 \caption{Experimental setup.  A wire is inserted into the
 laser cavity in order to generate a HG mode.  A mode converter
 consisting of two identical spherical (L$_{s}$) and cylindrical
 (L$_{c}$) lenses converts the HG mode to a LG mode of the same order. 
 The LG mode is then used to pump the BBO crystal, generating
 entangled photons incident on a HOM interferometer.  The beam
 splitter BS is mounted on a motorized stage.  Coincidence counts
 are recorded at detectors $D_{1}$ and $D_{2}$.}
\label{setup}
 \end{figure} 
%---------------------------END FIGURE------------------------------- 
The experimental setup, shown in Fig. \ref{setup}, consists of two
basic parts.  The first part is the generation of a Laguerre-Gaussian
(LG) mode using a mode converter, which transforms a Hermite-Gaussian
(HG) mode into a LG mode.  A detailed account of mode conversion can
be found in refs. \cite{beijersbergen93,padgett96}.  To create the
HG-mode, we insert a $25\,\mu$m diameter wire into the cavity of an
Argon laser, operating at $\sim 30\,$mW with wavelength $351.1\,$nm. 
The wire breaks the circular symmetry of the laser cavity. It is aligned in the horizontal or vertical and mounted on an
$xy$-translation stage.  By adjusting the position and orientation of
the wire, we can generate the modes HG$_{01}$, HG$_{10}$, HG$_{02}$, and HG$_{20}$.  The beam then passes through a mode converter consisting of two spherical lenses (L$_{s}$) with focal length $f_{s} = 500\,$mm and two cylindrical lenses (L$_{c}$) with focal
length $f{c} = 50.2\,$mm.  The first spherical lens is used for mode-matching and is located $\approx 1.88\,$m from the beam center of curvature.  The second spherical lens is placed confocal with the first, and is used to ``collimate" the beam.  The cylindrical lenses are placed ($d=f_{c}/\sqrt{2} \approx 35\,$mm) on either side of the focal point of lens L$_{s}$ and aligned
at $45^{\circ}$.  The cylindrical lenses transform the HG mode into a
LG mode of the same order by introducing a relative $\pi/2$ phase
between successive HG components (in the $\pm
45^{\circ}$ basis, due to the orientation of the cylindrical lenses) of
the input beam \cite{beijersbergen93,padgett96}.  The quality of the
output mode was checked by visual examination of the intensity profile
\cite{courtial99} as well as by interference techniques: using
additional beam splitters and mirrors (not shown), the interference of
the LG pump beam with a plane wave resulted in the usual spiral
interference pattern \cite{padgett96}.  \par The second part of the
setup is a typical HOM interferometer \cite{hong87}.  The Argon laser
is used to pump a $7\,$mm long BBO ($\beta$-BaB$_2$O$_4$) crystal cut
for type II phase matching, generating non-collinear entangled photons
by SPDC.  The down-converted
photons are reflected through a system of mirrors and incident on a
beam splitter with measured transmittance $T \approx 0.67$ and
reflectance $R \approx 0.33$.  Since the down-converted photons
are orthogonally polarized, a half-wave plate (HWP) is used to rotate the
polarization of one of the photons ($V \rightarrow H$). 
A computer-controlled stepper motor is used to adjust the position of
the beam splitter.  The detectors are EG \& G SPCM 200 photodetectors,
mounted on precision translation stages.  $D_{2}$ remained fixed while
a computer-controlled stepper motors were used to scan detector $D_1$
in the transverse plane.  Coincidence and single counts were
registered using a personal computer.  Interference filters ($1\,$nm
FWHM centered at $702\,$ nm) and $2\,$mm circular apertures where used
to align the HOM interferometer.  The transverse intensity profiles
were measured with the interference filters removed and circular
apertures with diameter $0.5\,$mm and $1\,$mm on $D_{1}$ and $D_{2}$,
respectively.

\section{Results and Discussion}
\begin{figure}
\includegraphics[width=8cm]{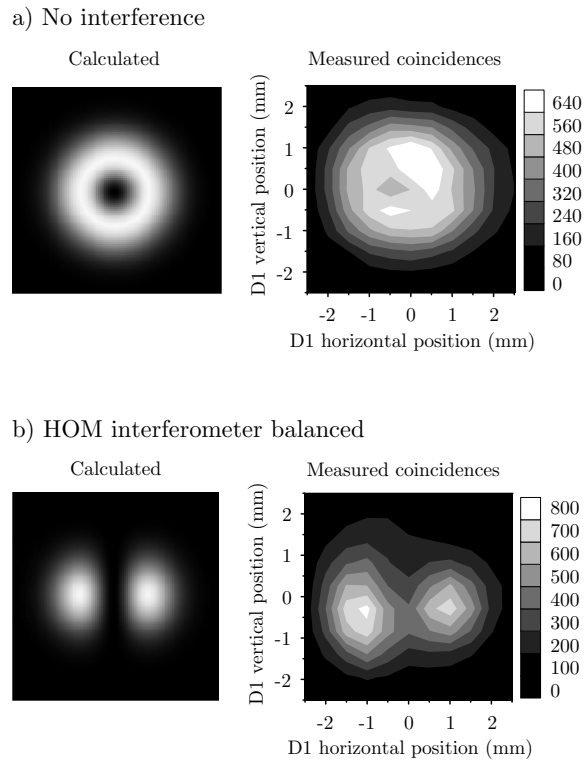}
\caption{Coincidence profiles predicted (left) and measured (right) when the crystal is pumped by a LG$_{0}^{1}$ beam. a) No-interference regime (Hong-Ou-Mandel interferometer unbalanced). b) Fourth-order interference regime (interferometer balanced).}
\label{res1}
\end{figure}

\begin{figure}
\includegraphics[width=8cm]{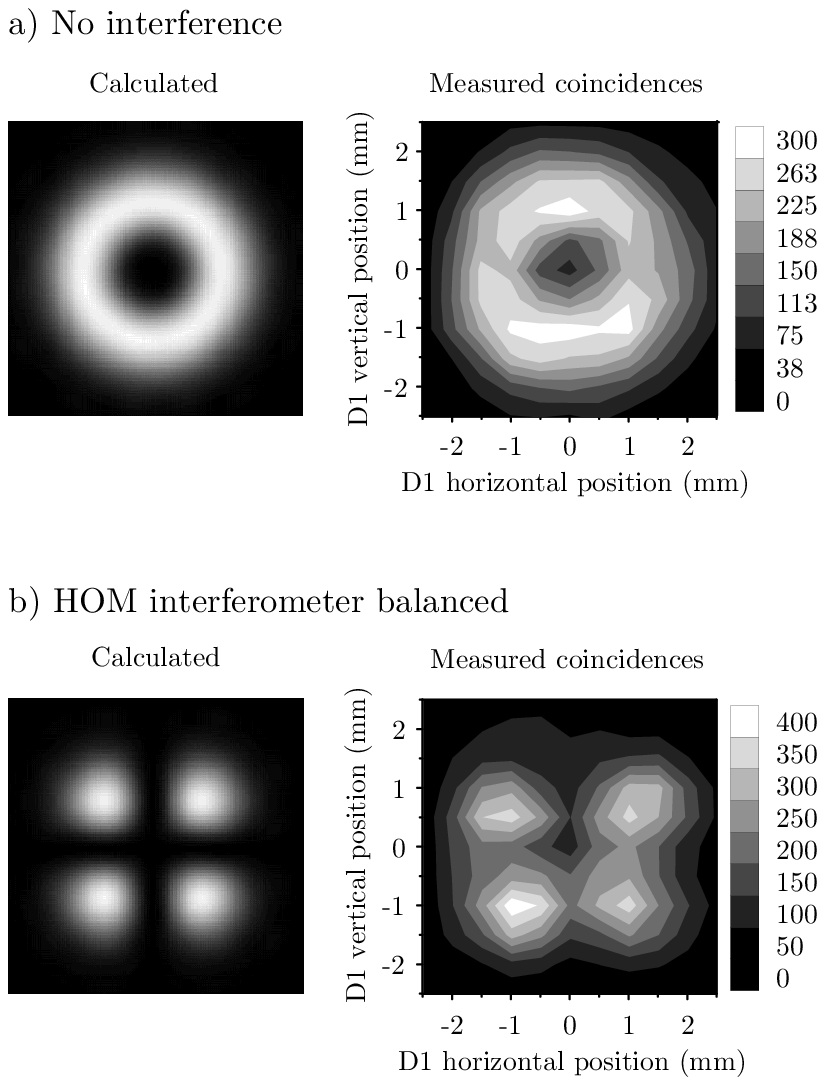} 
\caption{Coincidence profiles predicted (left) and measured (right) when the crystal is pumped by a LG$_{0}^{2}$ beam. a) No-interference regime (Hong-Ou-Mandel interferometer unbalanced). b) Fourth-order interference regime (interferometer balanced).}
\label{res2}
\end{figure}

\begin{figure}
\includegraphics[width=8cm]{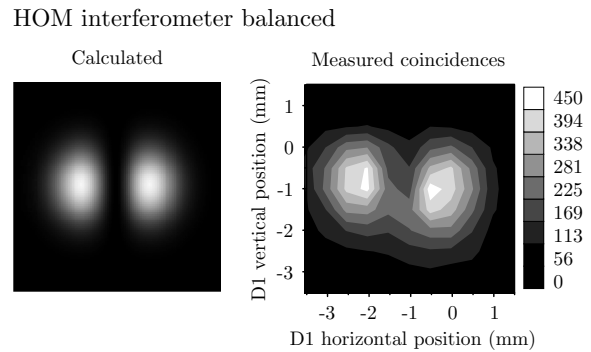} 
\caption{Coincidence profile predicted (left) and measured (right) when the crystal is pumped by a LG$_{0}^{1}$ beam, in the fourth-order interference regime (Hong-Ou-Mandel interferometer balanced). Detector $D_{2}$ was displaced by $\Delta x = \Delta y = 1$ mm.}
\label{res3}
\end{figure}

The results are shown in Figs. \ref{res1} to \ref{res3}. The left sides of the figures show the expected coincidence patterns, obtained from the squared modulus of Eq. (\ref{eq:psif2}) in the non-interfering regime (interferometer unbalanced), and from (\ref{eq:prob1}) in the fourth-order interference regime (interferometer balanced).  The right sides of the figures show the measured coincidences.

%---------------------------FIGURE-----------------------------------  
%---------------------------END FIGURE------------------------------- 
%---------------------------FIGURE-----------------------------------  
%---------------------------END FIGURE------------------------------- 
%---------------------------FIGURE----------------------------------- 
%---------------------------END FIGURE------------------------------- 

In Fig. \ref{res1}, the nonlinear crystal was pumped by a LG$_{0}^{1}$ ($l=1$) beam. Its intensity profile is shown in Fig. \ref{res1}a, in agreement with Eq. (\ref{eq:psif2}). In the interference regime, shown in Fig. \ref{res1}b, the two interference peaks predicted by Eq. (\ref{eq:prob1}) are easily seen. In Fig. \ref{res2}, the nonlinear crystal was pumped by a LG$_{0}^{2}$ ($l=2$) beam. Now, the interference pattern shows four peaks, in agreement with Eq. (\ref{eq:prob1}).

In order to test the the translational invariance of $\Psi(\bsy{\rho}_{1},\bsy{\rho}_{2})$, which leads to the conclusion that the two-photon state is entangled in orbital angular momentum, we repeated the measurement of Fig. \ref{res1}b, with detector $D_{2}$ displaced by $\Delta x = \Delta y =1$ mm. The interference pattern obtained is shown in Fig. \ref{res3}. The coincidence pattern measured by scanning $D_{1}$  is now dislocated by $\Delta x = \Delta y =-1$ mm, still in agreement with  Eq. (\ref{eq:prob1}).

\section{Conclusion}
We have shown experimentally that our theoretical description of the two-photon wave function is accurate. Information about its modulus and phase structure were obtained by direct coincidence detection and coincidence detection of fourth-order HOM interference effects, respectively.
The transfer of the plane wave spectrum of the pump beam to the fourth-order transverse spatial correlation function of the two-photon field generated by SPDC leads to the conservation and entanglement of the orbital angular momentum of the down-converted fields.
We should stress that this effect is restricted to the context of two approximations. The first is the paraxial approximation, in which our model for the transfer of plane wave spectrum in SPDC is based. However, the paraxial approximation is also the context in which the angular momentum carried by electromagnetic beams can be separated into an intrinsic part, associated to polarization, and an orbital part, associated to the transverse phase structure of the beam. The second approximation is the so-called thin crystal approximation. It is possible to show that this approximation would not be necessary if the non-linear medium were isotropic. The birefringence of the crystals used for SPDC causes non-conservations of the orbital angular momentum that are proportional to the crystal length. Rigorously speaking, orbital angular momentum would never be conserved in SPDC due to this effect. In thin crystals (few millimeters in length) however, it can be neglected. We believe that the arguments and the experiment reported here provide additional evidence of conservation and entanglement of the orbital angular momentum of light in SPDC, as well as the limits within which they should be understood.

\begin{acknowledgments}
The authors thank the Brazilian funding agencies CNPq and CAPES.
\end{acknowledgments}

%\bibliography{OAM-HOM}
\end{document}